\documentclass[twocolumn,showpacs,preprintnumbers,amsmath,amssymb,superscriptaddress]{revtex4}

\usepackage{graphicx,subfigure}
\usepackage{dcolumn}
\usepackage{bm}
\usepackage{psfrag}
\usepackage{multirow}

\def\Im {\mbox{Im}}
\def\be{\begin{equation}}       \def\ee{\end{equation}}
\def\bea{\begin{eqnarray}}      \def\eea{\end{eqnarray}}

\begin{document}
\title{The magnetoelectric coupling in the multiferroic compound $LiCu_{2}O_{2}$}
\author{Chen Fang} \affiliation{Department of Physics, Purdue
University, West Lafayette, IN 47907}\email{cfang@purdue.edu}\email{hu4@physics.purdue.edu}
\author{Trinanjan Datta}
\affiliation{Department of Chemistry and Physics, Augusta State
University, Augusta, GA 30904}\email{tdatta@aug.edu}
\affiliation{Department of Physics, Purdue University, West
Lafayette, IN 47907}
\author{Jiangping Hu}
\affiliation{Department of Physics, Purdue University, West
Lafayette, IN 47907}
\date{\today}
\begin{abstract}\label{abstract}
We investigate the possible types of coupling between
ferroelectricity and magnetism for the zig-zag spin chain multiferroic $LiCu_{2}O_{2}$  compound.
We construct a multi-order parameter phenomenological model for the 
material based on a group theoretical analysis. From our calculation
we conclude that a coupling involving the inter-chain magnetic
structure and ferroelectricity is necessary
to understand the experimental results of Park \emph{et.al.}
\cite{park}. Our proposed model is able to account for the electric
polarization flop in the presence of an externally applied magnetic
field. Furthermore, based on our theoretical model we can make specific
selection rule predictions about electromagnon  excitations present
in the $LiCu_{2}O_{2}$ system. We also predict that the electromagnon peaks measured in an $ac$-conductivity measurement
are field dependent.
\end{abstract}

\maketitle
\section{Introduction\label{sec:introduction}}
Cross coupling of magnetism and ferroelectricity in material
systems is an intriguing 
phenomenon. 
Presently there are chemical compounds in which both magnetism and ferroelectricity can exist
simultaneously. These novel systems are called multiferroics \cite{Kimura2003a,Hur2004a,Goto2004,Lottermoser2004,Lorenz2004,Higashiyama2004}.

Multiferroic materials have the unique feature that their magnetic properties
can be controlled by an electric field and the electrical properties by a magnetic field. Such a control 
allows for the possibility to fabricate multifunctional devices with great potential for technological applications in areas such
as memory, sensor, and
spintronic devices \cite{Hill, Fiebig, Khomskii,Cheong2007}.
The present 
interest in multiferroics 
arises from the controlled tunability of these systems. Examples of multiferroic systems
include the rare earth perovskites $RMnO_3$, $RMn_2O_5$ ($R$: rare earths - Gd, Dy, Tb...)
\cite{Kimura2003a,Kimura2003,Hur2004a,Kimura2005a, guha}, the
delafossite $CuFeO_2$ \cite{Kimura2006}, the spinel $CoCr_2O_4$
\cite{Yamasaki2006}, $MnWO_4$ \cite{Taniguchi2006}, $Ni_3V_2O_8$
\cite{Lawes2005}, the hexagonal ferrite $A_2Zn_2Fe_{12}O_{22} (A=
Ba,Sr)$ \cite{Kimura2005},  the zig-zag spin chain  compound
$LiCu_2O_2$ \cite{park} and so on.

The mixed valent zig-zag spin chain cuprate compound $LiCu_{2}O_{2}$ has been shown to be a multiferroic
material by Park \emph{et.al.} \cite{park}. 
The experimental data on the compound exhibits the canonical
multiferroic behavior of a $\pi /2$- flip in the direction of its
electrical polarization for a magnetic field perpendicular to the
axes of polarization. The data also reveals that the compound
exhibits a unique behavior of generating an electric polarization
parallel to the applied magnetic field direction. Due to
sample issues, the second experimentally observed feature has to be investigated
further before being confirmed experimentally \cite{comm}. Recently, a
theoretical attempt has been made to explain 
these behaviors for the $LiCu_{2}O_{2}$
compound using a stoichiometry argument \cite{moskvin-2008,drechsler-2008}. In this article, we will not provide any
physical explanation for the occurence of this effect within our proposed model and will focus on explaining the polarization flip only \cite{comm}.

In the microscopic theories \cite{Katsura2005, Hu2008, Sergienko2006} 
the   polarization  flip 
has hardly been studied so far. An explanation of the $\pi/2$- polarization flip in the conventional continuum theory \cite{mostovoy} requires 
an introduction of numerous 
anisotropic magnetic terms which is rather unnatural.   In this
paper we take a different approach using both a group theoretical
analysis involving the symmetry of the magnetic
structure \cite{Harris2005a} and a phenomenological multi-order
parameter magnetization model to study the experimental data
\cite{park}. The multi-order parameter
model is able to account for the electric polarization flip through  $\pi /2$- in the presence of an applied external magnetic field. We show
that a coupling involving the inter-chain magnetic structure and ferroelectricity  can be used
to understand the experimental results 
\cite{park} and in particular the polarization flip. The theory does not include 
any anisotropic terms.
Furthermore, the   presence of inter-chain coupling 
is supported by experimental evidence found in the Raman scattering experiment \cite{choi}.

The new inter-chain coupling term cannot be expressed in the
familiar $\vec{P}\cdot[(\vec M\cdot\vec\nabla)\vec M-(\vec\nabla\cdot\vec M)\vec M]$
form in the continuum limit. A hidden assumption behind the continuum coupling model
\cite{mostovoy} is that the magnetic order is described by a single order parameter.
When the lattice structure is complicated this assumption is not valid and the general
form of the magnetoelectric coupling can be different. The additional inter-chain
magnetoelectric coupling   derived for the $LiCu_2O_2$ compound 
is a reflection of the physics that there are novel magnetoelectric coupling terms in a multi-order magnetic structure.
Besides accounting for the observed experimental
features we also discuss the selection rules associated with the low
energy electromagnon excitations.

This paper is arranged in six sections. In section~\ref{subsec:crystalmagneticstructure}
we describe the crystal and magnetic structure of the 
$LiCu_{2}O_{2}$ compound under study. In
section~\ref{subsec:ramanscattering}  we discuss the experimental
evidence which indicates that inter-chain coupling has an important
role to play. In section~\ref{sec:symmetryoperations} we elucidate
the magnetic symmetry operations of the lattice and use group theory
to construct the possible intra- and inter- chain couplings present
in this compound. In section~\ref{sec:phenomodel} we propose the
phenomenological model   for the  $LiCu_{2}O_{2}$ system.   In
section~\ref{sec:hamiltonianselectionrules} we first state and then
analyze the theoretical model which explains the experimental data
of Park \emph{et.al.}\cite{park}. We then derive the selection rule
associated with the hybrid excitations of phonon and magnon termed
electromagnons.  Finally, in section~\ref{sec:summaryconclusion} we
provide a summary and state the main conclusions of our paper.

\section{The $LiCu_{2}O_{2}$ system \label{sec:experiments}}
\subsection{Crystal and magnetic structure\label{subsec:crystalmagneticstructure}}

\begin{figure}[!h]
\centering {
  \includegraphics[width=2.0in]{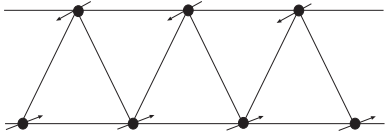}}
\caption{Spin chain: The multiferroic $LiCu_{2}O_{2}$ compound has pairs of $Cu^{2+}$ chains running parrallel to each other. Each chain is separated from the other by $b/2$ and forms the zig-zag triangular ladder structure shown in the figure. The black dots in the figure represent the magnetic $Cu^{2+}$ ions. These ions carry a spin-1/2 indicated by the arrow in the figure. The ladder is directed along the crystallographic $b$-axis.}
\label{fig:spinchain}
\end{figure}

The mixed valent cuprate compound $LiCu_{2}O_{2}$ is a ferroelectric material.
It crystallizes in an orthorhombic structure with the space group $Pnma$ (\# 62) \cite{Berger,Zvyagin,Roessli}.
The crystal structure of the compound can be visualized as follows. Consider two linear $Cu^{2+}$ chains 
propagating along the crystallographic $b$-axis. The two chains are displaced $b/2$ with respect to each
other and they form a zig-zag triangular ladder like structure as shown in Fig.~\ref{fig:spinchain}. The ladders are 
separated from each other by the non-magnetic $Li^{+}$ ions (distributed along the crystallographic $a$-axis) and by the layers
of nonmagnetic $Cu^{+}$ ions (distributed along the crystallographic $c$-axis). The magnetic behavior of the system is provided by the $Cu^{2+}$
ions which carry a spin-1/2. The letters $a,b,$ and $c$ represent the lattice constants of the simple orthorhombic crystal structure. 

The magnetic structure of $LiCu_{2}O_{2}$ has been determined from the
neutron scattering experiments \cite{masuda}. The magnetic
modulation vector obtained from these experiments is $\vec
{Q}=(0.5\frac{2\pi}{a},\frac{2\pi \xi}{b},0)$ where $\xi=0.174$ is the spiral modulation along the chain direction. The
letters $a,b,$ and $c$ represent the lattice constants of the simple
orthorhombic crystal structure as mentioned earlier.
Along the $a$-axis there is antiferromagnetic order and along the $b$-axis there is
a spiral order. 
Succesive spins on each rung are almost parallel to each other with each spin being rotated relative to each other by an angle
$\alpha=2\pi \xi$. Within each double chain any nearest-neighbor spins
from opposite legs are almost antiparallel and form an angle $\alpha
/2 =\pi \xi$. Neutron diffraction experiments
\cite{masuda} conclude that rotating spins lie in the $bc$ plane \cite{outofplane}.

\begin{center}
\begin{figure}[!t]
\centering \subfigure[Magnetic unit
cell]{\label{subfig:MagneticUnitCell}
  {\psfrag{a}{$a$}\psfrag{c}{$c$}\psfrag{1}{$S_{1}$}\psfrag{2}{$S_{2}$}\psfrag{3}{$S_{3}$}\psfrag{4}{$S_{4}$}
  \includegraphics[width=2.0in]{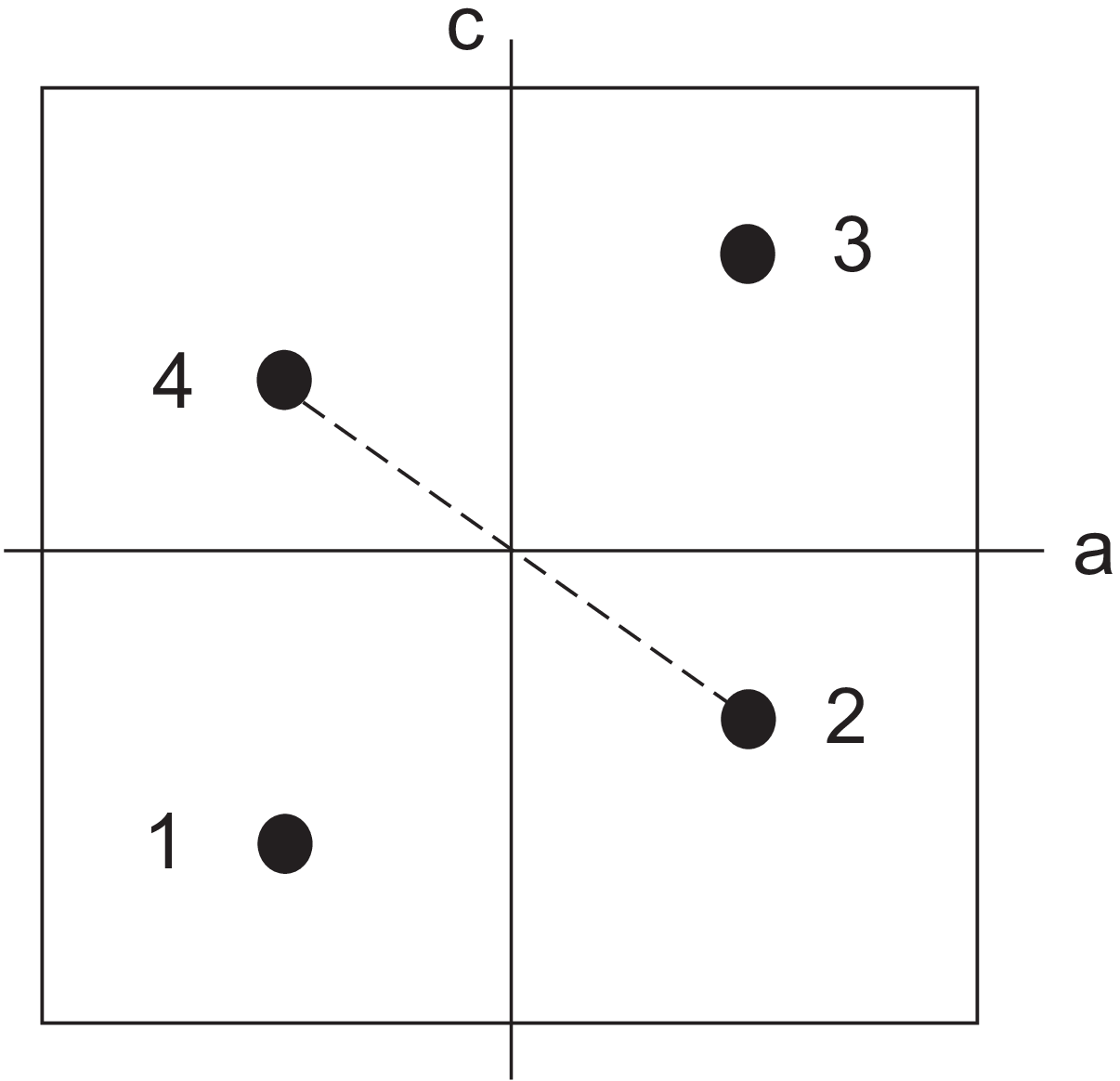}}}
  \subfigure[Side View of the unit cell]{\label{subfig:SideView}{\psfrag{b}{$b$}\psfrag{c}{$c$}\psfrag{B}{$b/2$}\psfrag{-B}{$-b/2$}
  \psfrag{1}{$S_{1}$}\psfrag{2}{$S_{2}$}\psfrag{3}{$S_{3}$}\psfrag{4}{$S_{4}$}
  \includegraphics[width=2.0in]{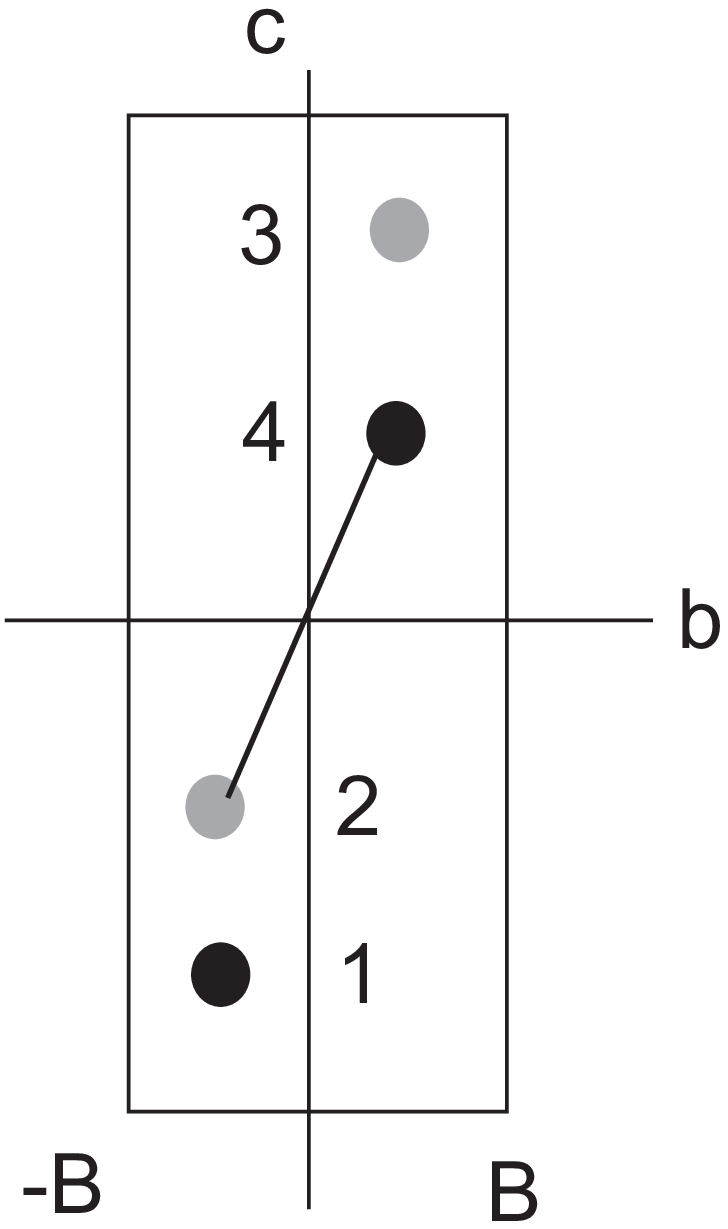}}}
\caption{The dashed and the bold line indicates the bond between the
$Cu^{2+}$ ions forming the zigzag chain.(a) Magnetic unit cell. The
four black dots represent $Cu^{2+}$ ions. (b) Side view of the same
magnetic unit cell. The center of inversion is situated halfway
between the two $Cu^{2+}$ ions. These ladders are separated from each other by the non-magnetic $Li^{+}$ ions (distributed along the crystallographic $a$-axis) and by the layers of nonmagnetic $Cu^{+}$ ions (distributed along the crystallographic $c$-axis).}\label{fig:magneticunitcell}
\end{figure}
\end{center}

\subsection{Experimental evidence of inter-chain coupling\label{subsec:ramanscattering}}

The present theoretical formulation of spiral ferroelectrics do not consider the role of inter-chain coupling.
For the $LiCu_{2}O_{2}$ compound, a spiral ferroelectric, there is experimental evidence 
from the Raman scattering experiment of Choi \emph{et.al.} \cite{choi} which suggests the presence of inter-chain
interactions. Referring to Fig.~1 of their paper one can clearly observe that similar peaks are observed
for both the parallel $(xx)$ and crossed polarizations $(xy)$. This indicates that inter-chain effects
are important and should be considered in the theoretical formulation of 
$LiCu_{2}O_{2}$. Similar behavior has also been observed in the Na-analog compound $NaCu_{2}O_{2}$ \cite{capogna,Gnezdilov}.
In this article we provide a theoretical justification using group theory for this important
experimental observation. We also construct a phenomenological theory considering
 the effects of inter-chain coupling in multiferroic compounds.


\section{Symmetry operations of the lattice and magnetic structure\label{sec:symmetryoperations}}

The goal of this section is to find a physically appropriate form
for the magnetoelectric coupling in $LiCu_{2}O_{2}$ subject to the
symmetry constraints 
  of  the magnetic structure. Physically
such a term should conserve translational symmetry, space inversion
symmetry, and time reversal symmetry. Phenomenological models 
\cite{mostovoy} are based on these considerations only. 
These models assume that a single unit cell of the lattice is a
point in the system without any structure.   An appropriate  
realistic model 
in these cases should include 
all the inequivalent atoms (considering both the type and location of the ion in space) in a single magnetic unit cell. 
This allows for the possibility to include more degrees of
freedom. For example, in the LiCu$_2$O$_2$ case there are
four spatially inequivalent magnetic ions in the magnetic unit
cell that we should consider. They are denoted as $\vec S_1$, $\vec
S_2$, $\vec S_3$, and $\vec S_4$ respectively (see
Fig.~\ref{fig:magneticunitcell}).

As the number of possible forms of the magnetoelectric coupling increases
rapidly with the degrees of freedom such a detailed model seems
too complicated to be useful. However there are symmetry
constraints which help to simplify the situation. These constraints
should be obeyed by any term in the Hamiltonian including the magnetoelectric coupling term. The terms should
be invariant under the lattice symmetry operations and time reversal symmetry. The lowest order magnetoelectric coupling which
preserves time reversal symmetry is a trilinear term. It is
linear in the electric polarization and bilinear in the magnetic
order parameters. To simplify the discussion, we consider the
coupling terms that involve the uniform spontaneous electric
polarization, $\vec P$, and not those including
the modulations of the polarization. The overall polarization
has been experimentally measured and confirmed. The general magnetoelectric coupling is
\begin{eqnarray}
\label{eq:hamiltonianME1}
H_{ME}=\lambda\sum_{\alpha\beta\gamma}P^\alpha\sum_{ij}\sum_{\vec
R\vec R'}C_{ij;\alpha\beta\gamma}(\vec R-\vec R')S^\beta_i(\vec
R)S^\gamma_j(\vec R')
\end{eqnarray}
where $\alpha,\;\beta,\;\gamma=x,\;y,\;z ;$ $i,\;j=1,\;2,\;3,\;4,$
and $\vec R$ denotes the position of the unit cell. The magnetic
structure in the compound  can be written as $S^\alpha_i(\vec
R)=S^\alpha_i(\vec Q)\exp(i\vec Q\cdot\vec R)+c.c$. Substituting
this expression 
into the above Hamiltonian we have
\begin{eqnarray}
\label{eq:hamiltonianME2}
H_{ME}=\lambda\sum_{\alpha\beta\gamma}P^\alpha\sum_{ij}(C_{ij;\alpha\beta\gamma}(\vec
Q)S^\beta_i(\vec Q)S^\gamma_j(-\vec Q)+c.c)
\end{eqnarray}
where $C_{ij;\alpha\beta\gamma}(\vec Q)$ is the Fourier transform of
$C_{ij;\alpha\beta\gamma}(\vec R-\vec R')$.

Not every term in the above magnetoelectric coupling conserves all
the symmetries. Only certain linear combinations do.  Specifically
we are interested in the lattice symmetry operations that conserve
the magnetic structure {with the magnetic propagation vector $\vec
Q$}. From group representation theory we know that these operations
constitute a subgroup of the full symmetry group of the lattice for
which we have one (and only one) two-dimensional representation,
$E$, $2_{b}$,$m_c$ and $m_c2_b$, all being two-by-two matrices .
The lattice symmetry operations which preserve the magnetic
structure and the magnetic propagation vector $\vec{Q}$ are listed
in Table~\ref{tab:table1}
\begin{table}[h]
\caption{Lattice symmetry operations which preserve the magnetic
structure and in turn the magnetic propagation vector- $\vec{Q}$. The
identity operation is represented by $E$. A two-fold rotation about
the crystallographic $b$-axis is represented by $2_{b}$. A
reflection about the $c$-axis is denoted by $m_{c}$ and finally a
combination of the the rotation-reflection by
$m_{c}2_{b}$.}
\begin{ruledtabular}
\label{tab:table1}
\begin{tabular}{l}
E${\bf r}=(x,y,z)$\\
$2_{b}{\bf r}=({\overline x},y+1/2,{\overline z})$\\
$m_{c}{\bf r}= (x-1/2, y, {\overline z} -1/2)$\\
$m_{c}2_{b}{\bf r} = ({\overline x}-1/2,y+1/2,z-1/2)$
\end{tabular}
\end{ruledtabular}
\end{table}

The two dimensional representations are given by
\begin{eqnarray}
\label{eq:symmetryoperations1}
E= \left( \begin{array}{cc}
1 & 0  \\
0 & 1 \end{array} \right); 2_{b}= \left( \begin{array}{cc}
-e^{-iqb/2} & 0  \\
0 &e^{-iqb/2} \end{array} \right)
\end{eqnarray}
\begin{eqnarray}
\label{eq:symmetryoperations2}
m_{c}= \left( \begin{array}{cc}
0 & -1  \\
1 & 0 \end{array} \right); m_{c}2_{b}= \left( \begin{array}{cc}
0 &-e^{-iqb/2}\\
-e^{-iqb/2} &0
 \end{array} \right)
\end{eqnarray}
For the above representation we can find  groups of symmetry adapted
variables which are transformed according to the representation
under the symmetry operations.  These variables can be constructed
as follows: a symmetry adapted variable is a column vector
(two-dimensional in this case) whose elements are a linear
combination of the original magnetic variables - $S_i^\alpha(\vec Q)$'s. The variables transform under
any operation of the subgroup in the same way they transform if left-multiplied by the corresponding matrix
for the same operation. These variables can be computed for the
$LiCu_2O_2$ case after deducing the transformation tables (see
Table~\ref{tab:table2} and Table~\ref{tab:table3}) of all the
members of the symmetry subgroup.
\begin{table}[h]
\caption{Symmetry operation: $2_{b}$, two-fold rotation about the crystallographic $b$-axis. In LiCu$_2$O$_2$ there are
four spatially inequivalent magnetic ions in the magnetic unit cell. They are denoted as $\vec S_1$, $\vec S_2$, $\vec S_3$, and $\vec S_4$ respectively (see Fig.~\ref{fig:magneticunitcell}). The subscripts $a,b,$ and $c$ denote the components along those crystallographic axis.}
\begin{ruledtabular}
\begin{tabular}{|c|c|c|c|}
\label{tab:table2}
$S_{1}$ & $S_{2}$ &$S_{3}$ &$S_{4}$ \\ \hline & & & \\
 $S^{'}_{1a}=-S_{3a}e^{-iq b/2}$ &$S^{'}_{2a}=-S_{4a}e^{-iq b/2}$
&$S^{'}_{3a}=-S_{1a}$ &$S^{'}_{4a}=-S_{2a}$ \\ \hline & & & \\
$S^{'}_{1b}=-S_{3b}e^{-iq b/2}$ &$S^{'}_{2b}=S_{4b}e^{-iq b/2}$
&$S^{'}_{3b}=S_{1b}$ &$S^{'}_{4b}=S_{2b}$ \\\hline & & & \\
$S^{'}_{1c}=-S_{3c}e^{-iq b/2}$ &$S^{'}_{2c}=-S_{4c}e^{-iq b/2}$
&$S^{'}_{3c}=-S_{1c}$ &$S^{'}_{4c}=-S_{2c}$
\end{tabular}
\end{ruledtabular}
\end{table}
\begin{table}[b]
\caption{Symmetry operation: $m_{c}$, reflection about the crystallographic $c$-axis. In LiCu$_2$O$_2$ there are
four spatially inequivalent magnetic ions in the magnetic unit cell. They are denoted as $\vec S_1$, $\vec S_2$, $\vec S_3$, and $\vec S_4$ respectively (see Fig.~\ref{fig:magneticunitcell}). The subscripts $a,b,$ and $c$ denote the components along those crystallographic axis.}
\begin{ruledtabular}
\begin{tabular}{|c|c|c|c|}
\label{tab:table3}
$S_{1}$ & $S_{2}$ &$S_{3}$ &$S_{4}$ \\ \hline & & & \\
$S^{'}_{1a}=-S_{2a}$ &$S^{'}_{2a}=S_{1a}$
&$S^{'}_{3a}=S_{4a}$ &$S^{'}_{4a}=-S_{3a}$ \\ \hline & & & \\
$S^{'}_{1b}=-S_{2b}$ &$S^{'}_{2b}=S_{1b}$
&$S^{'}_{3b}=S_{4b}$ &$S^{'}_{4b}=-S_{3b}$ \\\hline & & & \\
$S^{'}_{1c}=S_{2c}$ &$S^{'}_{2c}=-S_{1c}$ &$S^{'}_{3c}=-S_{4c}$
&$S^{'}_{4c}=-S_{3c}$
\end{tabular}
\end{ruledtabular}
\end{table}

As a result six groups of symmetry adapted variables are found,
that is, twelve elements in total. The number
of symmetry adapted vectors can also be deduced if we notice the fact that all of their elements should
 make a new basis   for  
the magnetic structure which requires the total number of elements
to be twelve,  the same as the original $S_i^\alpha$'s, and the
total number of vectors to be six. We   divide them into two sets as
listed below. The groups of symmetry adapted variables are
\begin{widetext}
\begin{eqnarray}
\label{eq:symmetryadaptedvariables1}
\mathcal{S}^{(1)}= \left\{\left( \begin{array}{c}
S_{1a}+e^{-iq b/2}S_{3a}   \\
S_{2a}-e^{-iq b/2}S_{4a} \end{array} \right),\left(
\begin{array}{c}
S_{1b}-e^{-iq b/2}S_{3b}   \\
S_{2b}+e^{-iq b/2}S_{4b} \end{array} \right),\left(
\begin{array}{c}
S_{2c}+e^{-iq b/2}S_{4c}   \\
S_{1c}-e^{-iq b/2}S_{3c} \end{array} \right)\right\}
\end{eqnarray}
\begin{eqnarray}
\label{eq:symmetryadaptedvariables2}
\mathcal{S}^{(2)}= \left\{\left( \begin{array}{c}
S_{2a}+e^{-iq b/2}S_{4a}   \\
-S_{1a}+e^{-iq b/2}S_{4a} \end{array} \right),\left(
\begin{array}{c}
S_{2b}-e^{-iq b/2}S_{4b}   \\
-S_{1b}-e^{-iq b/2}S_{3b} \end{array} \right),\left(
\begin{array}{c}
S_{1c}+e^{-iq b/2}S_{3c}   \\
-S_{2c}+e^{-iq b/2}S_{4c} \end{array} \right)\right\}
\end{eqnarray}
\end{widetext}
Using the properties of the symmetry adapted variables we
have
\begin{widetext}
\begin{eqnarray}
\label{eq:transformations1}
&\mathcal{S}^{(i)\dag}_\beta
M_{ij;\beta\gamma}\mathcal{S}^{(j)}_\gamma \stackrel{{2_b}}{\rightharpoonup} \mathcal{S}^{(i)\dag}_\beta(m_{2_b})^\dag
M_{ij;\beta\gamma}m_{2_b}
\mathcal{S}^{(j)}_\gamma=\mathcal{S}^{(i)\dag}_\beta
(\sigma_zM_{ij;\beta\gamma}\sigma_z)\mathcal{S}^{(j)}_\gamma\\
&\mathcal{S}^{(i)\dag}_\beta
M_{ij;\beta\gamma}
\mathcal{S}^{(j)}_\gamma \stackrel{{m_c}}{\rightharpoonup}\mathcal{S}^{(i)\dag}_\beta(m_{m_c})^\dag
M_{ij;\beta\gamma}m_{m_c}
\mathcal{S}^{(j)}_\gamma=\mathcal{S}^{(i)\dag}_\beta
(\sigma_yM_{ij;\beta\gamma}\sigma_y)\mathcal{S}^{(j)}_\gamma.
\end{eqnarray}
\end{widetext}
Hereafter we suppress the argument of $S$. We will consider $S$ to be $S(\vec
Q)$ and $S^+$ to be $S(-\vec Q)$. Having found the symmetry adapted
variables, we put the magnetoelectric coupling in the following
general { form} \bea \label{eq:hamiltonianME3}
H_{ME}=\lambda\sum_{\alpha}P^\alpha\sum_{i,j;\beta,\gamma}\mathcal{S}^{(i)\dag}_\beta
M_{ij;\beta\gamma}\mathcal{S}^{(j)}_\gamma+c.c.\eea where
$i,\;j=1,\;2$ and $M$ is an arbitrary two-by-twocoupling matrix. Using the properties of the symmetry
adapted variables we have
\begin{widetext}
\begin{eqnarray}
\label{eq:transformations2}
& \mathcal{S}^{(i)\dag}_\beta
M_{ij;\beta\gamma}\mathcal{S}^{(j)}_\gamma \stackrel{{2_b}}{\rightharpoonup}
\mathcal{S}^{(i)\dag}_\beta(m_{2_b})^\dag M_{ij;\beta\gamma}m_{2_b}
\mathcal{S}^{(j)}_\gamma=\mathcal{S}^{(i)\dag}_\beta
(\sigma_zM_{ij;\beta\gamma}\sigma_z)\mathcal{S}^{(j)}_\gamma\\
& \mathcal{S}^{(i)\dag}_\beta M_{ij;\beta\gamma}
\mathcal{S}^{(j)}_\gamma \stackrel{{m_c}}{\rightharpoonup}\mathcal{S}^{(i)\dag}_\beta(m_{m_c})^\dag
M_{ij;\beta\gamma}m_{m_c}
\mathcal{S}^{(j)}_\gamma=\mathcal{S}^{(i)\dag}_\beta
(\sigma_yM_{ij;\beta\gamma}\sigma_y)\mathcal{S}^{(j)}_\gamma
\end{eqnarray}
\end{widetext}
  Now we can apply the stated
constraints to find all the possible forms of the magnetoelectric
coupling.
\begin{enumerate}
\item \emph{Time reversal symmetry:} The expression above with its
tri-linearity automatically includes time reversal symmetry.

\item \emph{Lattice symmetry:} We focus on the subgroup of the lattice
symmetry operations. All three components of the electric
polarization behave differently under the symmetry operations. We
discuss each case separately. In the following we explicitly work
out the case for $P_c$ and simply list the result for the other two
components.  The symmetry properties of $P_c$ under the lattice
symmetry operations are $P_c \stackrel{m_c}{\rightarrow} -P_c$, $P_c
\stackrel{2_b}{\rightarrow}-P_c$,
$P_c\stackrel{I}{\rightarrow}-P_c$.  From the overall invariance of
the trilinear coupling we require that $M_{ij;\beta\gamma}$
anti-commute with both $\sigma_y$ and $\sigma_z$. From this we can
infer that $M_{ij}$ should be
proportional to $\sigma_x$ to 
{  preserve} the invariance under symmetry operations.  A similar
procedure can be applied to find the appropriate $M_{ij}$'s for
$P_a$ and $P_b$. The result can be summarized as follows: for $P_a$,
$M_{ij}\propto\sigma_y$; for $P_b$, $M_{ij}\propto1$; for $P_c$,
$M_{ij}\propto\sigma_x$. Later we will prove that the constant of
proportionality can be either real or purely imaginary based on
space inversion symmetry arguments.

\item \emph{Inversion symmetry:} Inversion operation is \emph{not} a member of the subgroup of the symmetry
operations that conserve the magnetic propagation vector $\vec Q$. Therefore it cannot be represented by a matrix acting on the
symmetry adapted variables. 
The mapping for space inversion operation is
\begin{eqnarray}
\label{eq:inversion}
&S_{1}^\alpha(q)\rightarrow
S_{3}^\alpha(-q),\;S_{2}^\alpha(q)\rightarrow S_{4}^\alpha(-q)\\
&S_{3}^\alpha(q)\rightarrow S_{1}^\alpha(-q),\;S_{4}^\alpha(q)\rightarrow S_{2}^\alpha(-q)
\end{eqnarray}
where $\alpha$ denotes the spin component. With this mapping we obtain the expression for the inversion operation in terms of the
symmetry adapted variables as
\begin{eqnarray}
\label{eq:inversiontansformation1}
I\cdot \mathcal{S}^{(i)}_{b}=\left( \begin{array}{cc}
-e^{-iqb/2} & 0  \\
0 &e^{-iqb/2}\end{array} \right)\mathcal{S}^{(i)*}_{b}
\end{eqnarray}
\begin{eqnarray}
\label{eq:inversiontansformation2}
I\cdot \mathcal{S}^{(i)}_{\alpha}=\left( \begin{array}{cc}
e^{-iqb/2} & 0  \\
0 &-e^{-iqb/2}\end{array} \right)\mathcal{S}^{(i)*}_{\alpha}
\end{eqnarray}
where $\alpha=a,c$ and $i=1,2$ are the indices for the symmetry adapted
variables. Below we explicitly work out the case for $P_c$ and list the result for the other two components of electric
polarization. We have
\begin{widetext}
\bea \label{eq:transformationpcandpa1}
P_c\mathcal{S}^{(i)\dag}_\beta\sigma_x\mathcal{S}^{(j)}_\gamma \stackrel{I}{\rightharpoonup}(-P_c)(\mathcal{S}^{(i)T}_\beta\sigma_z\sigma_x\sigma_z\mathcal{S}^{(j)*}_\gamma)=P_c(\mathcal{S}^{(i)\dag}_\beta\sigma_x\mathcal{S}^{(j)})^*,\eea
if $\beta,\;\gamma=a,\;c$ or $\beta=\gamma=b$, and \bea
\label{eq:transformationpcandpa2}
P_c\mathcal{S}^{(i)\dag}_\beta\sigma_x\mathcal{S}^{(j)}_\gamma\stackrel{I}{\rightharpoonup}(-P_c)(-\mathcal{S}^{(i)T}_\beta\sigma_z\sigma_x\sigma_z\mathcal{S}^{(j)*}_\gamma)=-P_c(\mathcal{S}^{(i)\dag}_\beta\sigma_x\mathcal{S}^{(j)}_\gamma)^*,\eea
\end{widetext}
if one and only one of $\beta$ and $\gamma$ is $b$. The electric
polarization and the total Hamiltonian must be real making the
proportionality constant in the first case 
real and in the second case purely imaginary. The complete result is summarized below. 
It shows that there are only six generic terms that do not violate any
of the symmetries of the system
\begin{eqnarray}
&P_a(\mathcal{S}^{(i)\dag}_\beta\sigma_y\mathcal{S}^{(j)}_\gamma+c.c)\label{eq:polarizationexpressions1a}\\
&P_b(\mathcal{S}^{(i)\dag}_\beta \mathcal{S}^{(j)}_\gamma+c.c)\label{eq:polarizationexpressions1b}\\
&P_c(\mathcal{S}^{(i)\dag}_\beta\sigma_{x} \mathcal{S}^{(j)}_\gamma+c.c)\label{eq:polarizationexpressions1c}
\end{eqnarray}
if $\beta,\gamma = a,c$ or $\beta = \gamma = b$, and \bea
&iP_a(\mathcal{S}^{(i)\dag}_\beta\sigma_y\mathcal{S}^{(j)}_\gamma-c.c)\label{eq:polarizationexpressions2a}\\
&iP_b(\mathcal{S}^{(i)\dag}_\beta \mathcal{S}^{(j)}_\gamma-c.c)\label{eq:polarizationexpressions2b}\\
&iP_c(\mathcal{S}^{(i)\dag}_\beta\sigma_{x}\mathcal{S}^{(j)}_\gamma-c.c)\label{eq:polarizationexpressions2c}
\eea
if one and only one of $\beta$ and $\gamma$ is $b$.
\end{enumerate}
To summarize our work up to this point we have exhausted all
possible forms of magnetoelectric coupling that are invariant under
(1) time reversal, (2) space inversion, and (3) all the lattice
symmetry operations that conserve the magnetic propagation vector by using
the symmetry adapted variables.


\section{Phenomenological magnetization model\label{sec:phenomodel}}

A simplified expression for the phenomenological model can be
obtained if we observe first that ferroelectricity coexists with
the non-collinear magnetic structure. This suggests that terms with
$\beta=\gamma$ can be excluded.
Second, the polarization along the $b$-axis is not observed. 
Therefore, the couplings with $P_b$ need not be considered. Third, the magnetic moments on the four
Cu$^{2+}$ atoms, assumed to be independent variables in our theoretical formulation,
form two zigzag chains \cite{Berger,Zvyagin,Roessli} extended in the
$b$-direction on each of which a spin density wave with a
propagation vector $q$ along the $b$-axis exists. To be specific, in
Fig.~\ref{fig:spinchain}, they propose that $S_2$ and $S_4$ are on
one zigzag chain and $S_1$ and $S_3$ in the adjacent
unit cell are on another zigzag chain. This observation 
drastically simplifies the expression of magnetoelectric coupling as
\bea
&S_1^\alpha=e^{-iqb/2}S_3^\alpha \label{eq:spinrelation1}\\
&S_2^\alpha=-e^{-iqb/2}S_4^\alpha \label{eq:spinrelation2} \eea
These relations reduce the symmetry adapted variables to their final form as displayed in Eq.~\ref{eq:reducedsymmetryadaptedvariables}.
Considering the fact that $S_{2a}=-e^{-iq b/2}S_{4a}$ and $S_{1b}=e^{-iq b/2}S_{3b}$ the
symmetry adapted variables $\mathcal{S}^{(1)}$ and
$\mathcal{S}^{(2)}$ become
\begin{eqnarray}
\label{eq:symmetryadapedvariables1} \mathcal{S}^{(1)}= \left\{\left(
\begin{array}{c}
2e^{-iq b/2}S_{3a}   \\
-2e^{-iq b/2}S_{4a} \end{array} \right),\left(
\begin{array}{c}
0   \\
0 \end{array} \right),\left(
\begin{array}{c}
0  \\
0 \end{array} \right)\right\} \nonumber \\
\end{eqnarray}
\begin{eqnarray}
\label{eq:symmetryadapedvariables2} \mathcal{S}^{(2)}= \left\{\left(
\begin{array}{c}
0  \\
0 \end{array} \right),\left(
\begin{array}{c}
-2e^{-iq b/2}S_{4b}   \\
-2e^{-iq b/2}S_{3b} \end{array} \right),\left(
\begin{array}{c}
2e^{-iq b/2}S_{3c}   \\
2e^{-iq b/2}S_{4c} \end{array} \right)\right\} \nonumber \\
\end{eqnarray}
All the magnetoelectric couplings terms { can be enumerated in terms} of { the} symmetry adapted variables by using the above two equations. After the insertion of Eqs.~\ref{eq:spinrelation1} and \ref{eq:spinrelation2} 
three of the six symmetry adapted variables reduce to zero.  We can
instead take the reduced symmetry adapted variable as \bea
\label{eq:reducedsymmetryadaptedvariables} \vec{\mathcal{S}}=
\left\{\left( \begin{array}{c}
S^{(1)}_a \\ -S^{(2)}_a \\
\end{array} \right),
\left( \begin{array}{c} -S^{(2)}_b \\   -S^{(1)}_b \\
\end{array}\right), \left(\begin{array}{c} S^{(1)}_c \\S^{(2)}_c
\\\end{array}\right)\right\} \eea
where the superfix denotes the chain number to which the spin belongs and the argument of the $S$'s are suppressed. However one should bear in mind that all the $S$'s are Fourier components and in general complex numbers. 

\subsection{Intra-chain interaction terms\label{subsec:intrachainterms}}
First let us enumerate all the intra-chain couplings that couple to
$P_c$ and $P_a$. Since $P_b$ has not been observed under any
conditions we ignore it in our theoretical consideration. These
couplings are \bea \label{eq:intrachainpcoupling}
iP_c(S^{(1)*}_bS^{(1)}_{c,a}-c.c.\pm(1)\rightarrow(2)) \eea and \bea
\label{eq:intrachainpacoupling}
P_a(S^{(1)*}_bS^{(1)}_{c,a}+c.c.\mp(1)\rightarrow(2)) \eea where the
upper sign in the $\pm$ ($\mp$) is for the case when the subscript
is $c$.

Our calculations show that an existing phenomenological theory
\cite{mostovoy} for a spiral ferroelectric material in which only
intra-chain interactions are included does not suitably explain the
present system of interest.
We demostrate through our calculations that some of the unexplained
experimental facts may be ascribed to inter-chain interactions
generated as a result of considering the details of the magnetic
unit cell. We also show that not all the intra-chain and the
inter-chain terms have the usual $\vec{P}\cdot[(\vec
M\cdot\vec\nabla)\vec M-(\vec\nabla\cdot\vec M)\vec M]$
phenomenological form.

In the previous theory it is assumed that the electric polarization
along the $c$-axis can be explained by its coupling with a cycloidal
spin density wave on the $bc$-plane. Referring to
Eq.~\ref{eq:polarizationexpressions1c} one finds that $P_c$ must be
coupled to
$S_{4b}S_{4c}^*+S_{3b}S_{3c}^*+c.c.$. As we know from above that $\vec S_4$ and $\vec S_3$ are the order parameters of different zigzag chains, we conclude that in this theory, the ferroelectric order $P_c$ couples to bilinear products inside a zigzag chain. Furthermore, when the magnetic field is applied along the $b$-axis, $P_a$ can also be induced, and it is explained by its coupling to a cycloidal spin density wave in the $ab$-plane. Using Eq.~\ref{eq:polarizationexpressions1a} we know that $P_a$ can only couple to $S_{4a}S_{4b}^*+S_{3a}S_{3b}^*+c.c.$ which is a product inside a single zigzag chain. Following this line of reasoning one easily sees that at least some unexplained multiferroic properties of the system must come from magnetoelectric couplings that involves more than one chain, that is, an inter-chain coupling term. 

The physical picture of \emph{ferroelectricity induced by twisted magnetism} has been widely recognized in multiferroic theories \cite{mostovoy}. { Through our calculation} we point out that this { physical} picture is based on a single chain one-dimensional physics 
and { has to modified to include} 
inter-chain interaction. This observation is based on symmetry
arguments and therefore promises to be independent of the specific
form of interaction.

In transforming the above expressions, which are the
Fourier components, to an expression in the real space, one should
notice that a imaginary unit $i$ corresponds to an odd number of
derivatives while the absence of an $i$ corresponds to an even number of
derivatives. Therefore, if we only consider the lowest order in the
continuous limit the above expressions can be written as
\bea \label{eq:intrachainpccouplingrealspace}
\lambda_{c}P_c(S^{(1)}_b(y)\partial_yS^{(1)}_{c,a}(y)-\partial_yS^{(1)}_b(y)S^{(1)}_{c,a}(y)
\nonumber \\\pm(1)\rightarrow(2))\nonumber\\
\eea and \bea \label{eq:intrachainpacouplingrealspace}
P_a(S^{(1)}_b(y)S^{(1)}_{c,a}(y)\mp(1)\rightarrow(2)) \eea
If the magnetic structure is a cycloidal spiral on the $bc$-plane we can easily see that $P_c\neq0$ which is in accord with the phenomenological theory.  The coupling in front of the $P_{c}$ { component of} polarization is dependent on the magnetic field along the $a$-axis and is sensitive to it. A discrepancy with the { intra-chain phenomenlogical} theory occurs when we consider the observed ferroelectricity flop from the $c$-axis to the the $a$-axis. In the phenomenological theory it is explained as derived from a spin flop from the cycloid on the $bc$-plane to the $ab$-plane. However if we 
consider a magnetic structure such as $\vec{S}(x)=S(\cos(qx),\sin(qx),0)$ 
we find that $P_a$ gives a modulated value and therefore averages to zero. 
Since the starting point from group theory is more general 
we believe that the phenomenological expression for the
magnetoelectric coupling \bea \vec{P}\cdot[(\vec
M\cdot\vec\nabla)\vec M-(\vec\nabla\cdot\vec M)\vec M] \eea does not
capture the entire physical picture. Specifically in this case the
$P_c[(\vec M\cdot\vec\nabla)\vec M-(\vec\nabla\cdot\vec M)\vec M]_c$
has the correct form while the expression for $P_a[(\vec
M\cdot\vec\nabla)\vec M-(\vec\nabla\cdot\vec M)\vec M]_a$ is invalid
atleast for the $LiCu_{2}O_{2}$ case.
The derivation of the phenomenological theory \cite{mostovoy} 
assumes certain general symmetries of the Hamiltonian to construct the 
magnetoelectric coupling for $P_a$ and $P_c$. Based on such an
argument there is no reason for the coupling to be different.

The starting point of the phenomenological theory \cite{mostovoy}
considers each unit cell as a single site with magnetization $M$. This assumption does not hold  if the little group of the lattice
symmetry operations has no one-dimensional representation.
If the system can be effectively described by only the unit cell
magnetization $M$, then all the little group operations should
commute
and there must exist at least one one-dimensional representation. In the case of 
multiferroics such as $RMnO_3$ \cite{Kimura2003a,Kimura2003} and $Ni_3V_2O_8$ \cite{Lawes2005} there are
one-dimensional representations and the phenomenological theory
provides a correct description \cite{Harris2005a}. For other multiferroic systems such
as $R$Mn$_2$O$_5$ and $LiCu_2$O$_2$ only two-dimensional
representations exist because of the antiferromagnetic structure
along the $a$-axis. As a result the conventional phenomenological theory doesnot apply.
From the above discussion we clearly see that the intrinsic structure of the unit cell is important 
and therefore must be taken into consideration. 

\subsection{Inter chain interactions\label{subsec:interchainterms}}
{  From the expressions for the} intra-chain terms
we can conclude that 
an explanation {  for the} net magnetization along the $a$-axis
{  cannot be obtained within the intra-chain theory.} 
We therefore go {  beyond the scope of that theory and include the effects of} inter-chain coupling. Since there are several terms involved we 
first focus on those that couple to $P_c$. These expressions are
\bea \label{eq:interchainpccoupling}
P_{c}(\vec{S^{(1)}}\times\vec{S^{(2)}})_{b}\\P_c(\lambda_aS^{(1)}_aS^{(2)}_a+\lambda_bS^{(1)}_bS^{(2)}_b+\lambda_cS^{(1)}_cS^{(2)}_{b})
\eea where the $\lambda$'s are arbitrary constants. Physically the
first term could induce a polarization when the spins are aligned on
the $ac$-plane and $\vec{S^{(1)}}\bot\vec{S^{(2)}}$. If the
$\lambda$'s are close to each other the second term can be viewed as
$\lambda P_c(\vec{S}_1 \cdot \vec{S}_2)$. This term is minimized
with a collinear spin arrangement. The terms that couple to $P_{a}$
are
\begin{widetext}
\bea \label{eq:interchainpacoupling}
P_{a}(\vec{S^{(1)}}\times\partial_y\vec{S^{(2)}}+\vec{S^{(2)}}\times\partial_y\vec{S^{(1)}})_{b}
\\P_a(\beta_aS^{(1)}_a\partial_yS^{(2)}_a -\beta_a\partial_yS^{(1)}_aS^{(2)}_a +\beta_bS^{(1)}_b\partial_yS^{(2)}_b-\beta_b\partial_yS^{(1)}_bS^{(2)}_b+\beta_cS^{(1)}_c\partial_yS^{(2)}_c-\beta_c\partial_yS^{(1)}_cS^{(2)}_c)
\eea
\end{widetext}
Same as above we can see that the first term induces a nonzero
ferroelectricity with the alignment on the $ac$-plane and the
collinearity between the two chains while the second term is
minimized when $\vec{S^{(1)}}\bot\vec{S^{(2)}}$. The above group
theory analysis yields the possible intra-chain and inter-chain
magnetoelectric terms. The coupling parameters in the interactions
cannot be determined without a proper understanding of the
microscopic theory of the material compound and without improved
experimental results.

\section{Effective Hamiltonian and electromagnon selection rules\label{sec:hamiltonianselectionrules}}
In this section we conjecture a simple effective Hamiltonian which
can explain the basic phenomena observed in the $LiCu_2O_2$
experiments \cite{park}. In particular we provide an explanation for
the polarization flip in the
presence of an external magnetic field. 
Furthermore, using the phenomenological Hamiltonian 
we also  
predict a  
selection rule to be obeyed by the hybrid excitations of phonon and
magnon termed as electromagnons \cite{Sushkov2007} in the
literature. It is hoped that these electromagnon excitations can be
experimentally detected in future experiments.

\subsection{Effective Hamiltonian\label{subsec:effectivehamiltonian}}
The 
effective Hamiltonian that we propose includes only two terms.
One term involves the intra-chain coupling 
which is responsible for the electric polarization along the
$c$-axis. The other term involves the inter-chain coupling that is
responsible for the electric
polarization along the $a$-axis. 
The magnetoelectric coupling Hamiltonian, $H_{em}$, is given by \bea
\label{eq:emhamiltonian}& & H_{em}= \lambda_c P_c(\vec
S^{(1)}\times\partial_y\vec S^{(1)}+\vec S^{(2)}\times\partial_y\vec
S^{(2)})_a \label{eq:polarc}\nonumber \\ & & +\lambda_a P_a(\vec
S^{(1)}\times\partial_y\vec S^{(2)}+\vec S^{(2)}\times\partial_y\vec
S^{(1)})_b\label{eq:polara} \eea  where the symbols have the same
meaning as before. Both these terms satisfy the underlying lattice
and magnetic symmetry requirements.

The proposed effective Hamiltonian explains the basic physics  of
the $LiCu_{2}O_{2}$ system. First, 
in the absence of any external magnetic field according to the
neutron scattering experiments \cite{masuda} the spins lie on the
$bc$-plane.  We have for the ground state spin configuration,
$\vec{S}^{(1)}_{o}$ and $\vec{S}^{(2)}_{o}$, in the two chains \bea
\label{eq:spincomponentsbcplane} \vec
S^{(1)}_{o}=S(0,\cos(qy),\sin(qy))\\ \vec
S^{(2)}_{o}=S(0,\cos(qy+\delta\phi),\sin(qy+\delta\phi) \eea  where
$\delta\phi$ is the phase difference between the spin configurations
of the two chains.  Based on Eq.~\ref{eq:emhamiltonian} this spin
configuration leads to an electric polarization along the $c$-axis
\begin{eqnarray}
 P_c\propto(\vec S^{(1)}\times\partial_y\vec S^{(1)}+\vec
S^{(2)}\times\partial_y\vec S^{(2)})_a = const.
\end{eqnarray}
Second, in the presence of an applied magnetic field along the
$b$-axis
the spin should flip to the $ac$ plane when the 
field is larger than a certain critical value. This spin-flop
transition is expected in a very general magnetic model with
relatively isotropic couplings and spiral magnetism. This is 
{  similar to} the spin-flop transition in the antiferromagnetic
Heisenberg model in the presence of a Zeeman magnetic field.
Therefore the expected ground state
spin configuration in the two chains are 
\bea \label{eq:spincomponentsacplane} \vec
S^{(1)}_{o}=S(\cos(qy),0,\sin(qy))\\ \vec
S^{(2)}_{o}=S(\cos(qy+\delta\phi,0,\sin(qy+\delta\phi)) \eea
Based on Eq.~\ref{eq:emhamiltonian} this spin 
arrangement leads to an electric polarization along the $a$-axis
when the spin structure in both the spin chains are in phase, that
is, $\delta\phi=0$
\begin{eqnarray}
 P_a\propto(\vec S^{(1)}\times\partial_y\vec S^{(2)}+\vec
S^{(2)}\times\partial_y\vec S^{(1)})_b = const.
\end{eqnarray}

The theoretical model developed above suggests that the polarization
in the two crystallographic
direction comes from two different types of coupling. The polarization along the $c$-axis arises  
from the intra-chain coupling while the polarization along the
$a$-axis stems from the
inter-chain coupling. Furthermore, in formulating  this model we do not 
include strong anisotropy in the spin model to explain the flop
transition of the electric polarization in the presence of an
applied external magnetic field.

The proposed Hamiltonian 
explains the polarization flop by an applied magnetic field without 
introducing complicated anisotropic magnetic couplings
\cite{mostovoy}. The above model predicts explicit magnetic
configurations corresponding to the polarization
directions which can be 
verified in future experiments.

\subsection{Electromagnon selection rule\label{subsec:electromagnons}}
Electromagnons are a combination of phonon and magnon excitations \cite{Sushkov2007}. From the phenomenological model we can 
derive an explicit selection rule governing the electromagnons. The
selection rule can be stated as follows: Electromagnetic waves that
are polarized perpendicular to the bulk electric polarization can be
absorbed.
To be more specific 
in zero magnetic field only those electromagnetic waves that are polarized along the crystallographic $a$-axis can couple to the magnons. 
However, with an applied magnetic field along the $b$-axis only those waves polarized 
in the crystallographic $c$-axis can couple to the magnons.

The selection rules can be obtained in the following manner. We
first consider the case with the external magnetic field absent.
The dynamics of the system can be derived by assuming a small
deviation in the ground state properties \bea
\label{eq:smalldeviations} \vec P=\vec P_{o}+\vec u\\ \vec
S^{(1)}=\vec S^{(1)}_{o}+\delta\vec S^{(1)} \\ \vec S^{(2)}=\vec
S^{(2)}_{o}+\delta\vec S^{(2)} \eea where $\vec P_{o}$,
$\vec{S}^{(1)}_{o}$, and $\vec{S}^{(2)}_{o}$ are the ground state
values and the small deviations are indicated by $\vec u$,
$\delta\vec S^{(1)}$, and $\delta\vec S^{(2)}$. Energy minimization
requires that the first order terms vanish in the ground state. We
therefore study the second order terms in the perturbation expansion
to see how the dynamical degrees of freedom are coupled. Also the
spin being a length-preserving vector we know that
$\delta\vec S_i\bot\vec S_{i0}$. 
 Defining $\vec n_1=(0,\cos(qy),\sin(qy)),\;\vec n_2=(0,-\sin(qy),\cos(qy))$, we
have $\delta\vec S_1=m^{(1)}_1\vec n_2+m^{(1)}_2 \vec a$ (where
$\vec a$ is the unit vector along the $a$-axis). We now consider the
second order coupling that involves $\vec u_{c}$ and $\delta\vec
S_1$ derived from Eq.~\ref{eq:polarc}

\bea \label{eq:selectionrulecouplingpc} &u_c(\delta\vec
S^{(1)}\times\partial_y\vec S^{(1)}_{o}- \partial_y\delta\vec
S^{(1)}\times\vec S^{(1)}_{o})_a\nonumber \\&=S^2u_c(q(m^{(1)}_1\vec
n_2+m^{(1)}_2 \vec a)\times\vec n_2\nonumber \\
&-(\partial_ym^{(1)}_1\vec n_2-qm^{(1)}_1\vec n_1+\partial_y\vec
a)\times\vec n_1)_a\nonumber \\&=S^2\partial_ym_{1}u_c \eea In
momentum space the above expression can be written as \bea
\label{eq:s2couplingmomentumspace} S^2u_c(-\vec
k)(-ik_y)m^{(1)}_1(\vec k) \eea From this we infer that the zero
mode phonon $u_c(\vec k=0)$ does not couple to the magnons.
Therefore, if we compute the optical conductivity $\Im G_{uu}(\vec
k=0,\omega)$ in the lowest order (second order) the coupling does
not contribute. This shows that when $P_c \neq 0$ the phonons along
the same direction do not couple to the magnons. These phonons make
no contribution to the electromagnons that may be observed in an
optical conductivity measurement. However, the higher order terms
are still coupled. A similar type of result can be obtained with the
terms involving $\delta\vec S_2$. This verifies the first part of
our selection rule by showing that when an electric polarization is
present along the $c$-axis only the phonons
polarized along the $a$-axis can couple to the magnons 
to generate the electromagnons.

The coupling between the phonons polarized along the $a$-axis and
the magnons can be derived as follows. For $u_a$, from
Eq.~\ref{eq:polara} we have \bea \label{eq:selectionrulecouplingpa}
&u_a(\delta\vec S_1\times\partial_y\vec S_{20}-\vec
S_{20}\times\partial_y\delta\vec S_{1})_b\nonumber
\\&=S^2u_a(-q\cos(qy)m^{(1)}_2+\sin(qy)\partial_ym^{(1)}_2) \eea
Therefore the total effective Hamiltonian, $H^{eff}_{d}$, 
describing the dynamics of the phonon and the magnon can be written
as
\begin{eqnarray}\label{eq:effectivedynamichamiltonian}
& &H^{eff}_{d}= \sum_{p}\{{h\omega_0}b^\dag(p) b(p)
+\frac{\rho p^2}{2} \sum_i  m_2^{\dag(i)}(p)m_2^{(i)}(p) +\nonumber \\
& &
\frac{\lambda_aS^2}{2}(b^\dag(p+q)+b^\dag(p-q))[-qm_+(p)-pm_+(p)]\}
\end{eqnarray}
where $h$ is the Planck's constant, $\rho\sim J $ (the effective
magnetic coupling strength in the magnetic chain), $q$ is the
incommensurate wavevector, $\omega_0$ is the bare phonon energy,
$b_p$ is phonon annihilation operator, and $m_+ =\sum_i m^{(i)}_2$.

A qualitative understanding of the electromagnon frequency for the $LiCu_{2}O_{2}$ compound can be obtained by using a spinwave analysis. The analysis was done for the case when the optical phonon frequency is much larger than the magnon frequency of interest. This implies that if we measure the ac-conductivity versus frequency one will observe a peak at the frequency of the magnon. We should also note that two points justify the neglect of the dynamic lattice degrees of freedom, the displacement field $u$, for the $LiCu_{2}O_{2}$ system. First, the optical phonon frequency is much higher than the magnon frequency. Second, the magnetoelectric coupling is very small compared with to the other multiferroics, like the $113$ systems, making it unnecessary to explicitly include the dynamic degrees of freedom in the dielectric displacement \cite{Katsura2005}. We now perform a standard spinwave analysis \cite{Katsura2005} about the ground state to find the frequency at the desired wavevector, $\vec k=\vec Q$.

The model Hamiltonian for $LiCu_{2}O_{2}$ proposed by \cite{masuda} is
\begin{eqnarray}
H=\sum_{i,j}(J_{1}\vec{S}_{i,j}\cdot
\vec{S}_{i+1,j}+J_{2}\vec{S}_{i,j}\cdot \vec{S}_{i+2,j}\nonumber
\\+J_{4}\vec{S}_{i,j}\cdot \vec{S}_{i+4,j}
+J_{\perp}\vec{S}_{i,j}\cdot \vec{S}_{i,j+1})+DS_\perp^2.
\end{eqnarray}
Here we have included an easy plane anisotropy which generally exists due to the anisotropy
of the lattice and this term favors the easy plane of spin
alignment in the ground state, say, the bc-plane in zero field,
but the ac-plane in a strong magnetic field along the b-axis. We
solve for the spin wave dispersion. We employ the rotating frame
of reference coordinate system to write the spin vector at any
site relative to the other rotating one as
\begin{eqnarray}
\vec{S}_{i,j}= S^{\xi}_{i,j}\vec{e}_{x}+[S^{\eta}_{i,j}\cos(\vec{Q}\cdot\vec{R}_{i,j})+ S^{\zeta}_{i,j}\sin(\vec{Q}\cdot\vec{R}_{i,j})]\vec{e}_{y}\nonumber\\+[-S^{\eta}_{i,j}\sin(\vec{Q}\cdot\vec{R}_{i,j})+ S^{\zeta}_{i,j}\cos(\vec{Q}\cdot\vec{R}_{i,j})]\vec{e}_{z}\nonumber \\
\end{eqnarray}
In the above equation the magnetic propagation vector is given by $\vec{Q}=(\frac{\pi}{a},\frac{2\pi \xi}{b},0)$ where $\xi=0.174$ is the
spiral modulation along the chain. Inserting the above form for the spin vector we recast the Hamiltonian in terms of $S^{\xi},S^{\eta}$ and $S^{\zeta}$ components. We then compute the effective magnetic field components using the formula $\vec{H}_{eff}=-\vec{\nabla}_{\vec{S}_{i,j}}H$. We then solve for the equation of motion for the spin components using the effective magnetic field equations and spin components in terms of the Holstein-Primakoff formalism listed below
\begin{eqnarray}
S^{\xi}_{m,n}=\sqrt{\frac{S}{2}}(a_{m,n}+a^{*}_{m,n})\\
S^{\eta}_{m,n}=-i\sqrt{\frac{S}{2}}(a_{m,n}-a^{*}_{m,n})
\end{eqnarray}
The equations of motion are given by
\begin{eqnarray}
\hbar {\stackrel{.}{S}^{\xi}_{m,n}}=S^{\eta}_{m,n}H^{\zeta}_{eff} - S^{\zeta}_{m,n}H^{\eta}_{eff}\\
\hbar {\stackrel{.}{S}^{\eta}_{m,n}}=S^{\zeta}_{m,n}H^{\xi}_{eff} - S^{\xi}_{m,n}H^{\zeta}_{eff}
\end{eqnarray}
where in the above equation we set $S^{\zeta}_{i,j}=S$ since it is the direction in which the spin average points. Using the equations for the effective magnetic field, the spin components, and finally Fourier transforming we obtain the dispersion as

\bea -\frac{\hbar}{2S}\dot S^{\xi}(\vec q)=S^{\eta}(\vec
q)(J(\vec Q)-\frac{J(\vec Q+\vec q)+J(\vec Q-\vec
q)}{2}),\eea\bea-\frac{\hbar}{2S}\dot S^{\eta}(\vec q)=S^{\xi}(\vec
q)(J(\vec q)-J(\vec Q)+D),\eea where we define $J(\vec
q)=J_1\cos(q_{b}b)+J_2\cos(2q_{b}b)+J_4\cos(4q_{b}b)+J_\perp\cos(q_{a}a)$ where $q_{a}$ and $q_{b}$ refer to the $a$ and $b$ components
of the wavevector. From the equations above it is obvious that at $\vec q=\vec Q$ we have
\bea\hbar\omega(\vec Q)=2S\sqrt{(\frac{J(2\vec Q)+J(0)}{2}-J(\vec Q))D}.\eea

The electromagnon dispersion indicates that the frequency is proportional to the square root of the easy plane anisotropy, $D$. In zero field, the easy plane is the $bc$-plane. With an applied magnetic field along the $a$-axis, this field will effectively increase the anisotropy in the form of $D\sim D_0+H_{a}^2/2 \rho$ \cite{chen:epl}, where $\rho$ is the spin stiffness. This should lead to a hardening of the electromagnon frequency (a right shift of the peak in the ac-conductivity measurement). If the magnetic field is applied along the $b$-axis, it effectively diminshes the easy plane anistropy in the form of $D \sim D_0 -H_{b}^2/2\rho$. Therefore, one must observe the softening along with the increase of the magnetic field and at the point where the frequency becomes zero, the magnon mode becomes unstable and the spin-flop transition happens 

The above mechanism explains the observed phase transition at a certain magnetic field along the $b$-axis as the destabilization of one electromagnon mode. In the high field phase, the spins are lying in the $ac$-plane, making the $ac$-plane an easy plane described by an anisotropy term. In this phase, further increasing the field along the $b$-axis will increase the easy plane anisotropy, whereas applying a field along the $a$-axis will decrease the anisotropy. Therefore, we predict that the electromagnon hardens with an increase of magnetic field along the $b$-axis, and softens with an increase of magnetic field along the $a$-axis. This is opposite to what we should see in the low field phase.

\section{Summary and Conclusions\label{sec:summaryconclusion}}
In this paper we 
analyze the possible types of magnetoelectric coupling in the
recently studied multiferroic compound $LiCu_{2}O_{2}$.
Based on a group theoretical 
analysis we construct a multi-order parameter phenomenological model
for the double chain zig-zag structure. We show that a coupling
involving the inter-chain magnetic structure and ferroelectricity
is crucial 
in understanding the results of Park \emph{et.al.} \cite{park}. 
This constructed model for the multiferroic $LiCu_{2}O_{2}$
compound can
explain the polarization flop from the $c$-axis to the $a$-axis with
the applied magnetic field along the $b$-axis.
The model also makes specific selection rule predictions about the
hybrid phonon and magnon excitations called electromagnons.
We also predict that 
the electromagnon peaks measured in an $ac$-conductivity measurement
are field dependent and behave in opposite ways in the $P//a$
phase and $P//c$ phase. However, since the value of the
polarization in this material is rather weak it will require a
very high resolution spectroscopy measurement to observe the
electromagnons in the actual system.

The results in this paper 
suggest that the effective theory of magnetoelectric coupling is
richer in a system described by multi-order parameters where several 
magnetoelectric coupling terms can be constructed, than that in a system described by a single
order parameter. Similar physics has also been demonstrated 
for the $RMn_2O_5$\cite{chen:epl} multiferroic system and it should be thoroughly investigated for the other materials.

The model we propose in this paper could be oversimplified. However,
at present only a limited  set of experimental results  are
available for this compound. We believe this  phenomenological model
is a first step towards understanding the unique and novel
magnetoelectric coupling observed in the $LiCu_{2}O_{2}$ compound.
It is open to future experiments to determine the relevance of the
other magnetoelectric coupling terms which we
derived in this paper based on a group theoretical calculation. 

\section{Acknowledgements}
JP Hu acknowledges the extremely useful discussions with Prof. S.W.
Cheong.
\bibliography{LiCuObib}

\begin{thebibliography}{38}
\expandafter\ifx\csname natexlab\endcsname\relax\def\natexlab#1{#1}\fi
\expandafter\ifx\csname bibnamefont\endcsname\relax
  \def\bibnamefont#1{#1}\fi
\expandafter\ifx\csname bibfnamefont\endcsname\relax
  \def\bibfnamefont#1{#1}\fi
\expandafter\ifx\csname citenamefont\endcsname\relax
  \def\citenamefont#1{#1}\fi
\expandafter\ifx\csname url\endcsname\relax
  \def\url#1{\texttt{#1}}\fi
\expandafter\ifx\csname urlprefix\endcsname\relax\def\urlprefix{URL }\fi
\providecommand{\bibinfo}[2]{#2}
\providecommand{\eprint}[2][]{\url{#2}}

\bibitem[{\citenamefont{Park et~al.}(2007)\citenamefont{Park, Choi, Zhang, and
  Cheong}}]{park}
\bibinfo{author}{\bibfnamefont{S.}~\bibnamefont{Park}},
  \bibinfo{author}{\bibfnamefont{Y.~J.} \bibnamefont{Choi}},
  \bibinfo{author}{\bibfnamefont{C.~L.} \bibnamefont{Zhang}}, \bibnamefont{and}
  \bibinfo{author}{\bibfnamefont{S.-W.} \bibnamefont{Cheong}},
  \bibinfo{journal}{Phys. Rev. Lett.} \textbf{\bibinfo{volume}{98}},
  \bibinfo{pages}{057601} (\bibinfo{year}{2007}).

\bibitem[{\citenamefont{Kimura et~al.}(2003{\natexlab{a}})\citenamefont{Kimura,
  Goto, Shintani, Ishizaka, Arima, and Tokura}}]{Kimura2003a}
\bibinfo{author}{\bibfnamefont{T.}~\bibnamefont{Kimura}},
  \bibinfo{author}{\bibfnamefont{T.}~\bibnamefont{Goto}},
  \bibinfo{author}{\bibfnamefont{H.}~\bibnamefont{Shintani}},
  \bibinfo{author}{\bibfnamefont{K.}~\bibnamefont{Ishizaka}},
  \bibinfo{author}{\bibfnamefont{T.}~\bibnamefont{Arima}}, \bibnamefont{and}
  \bibinfo{author}{\bibfnamefont{Y.}~\bibnamefont{Tokura}},
  \bibinfo{journal}{Nature} \textbf{\bibinfo{volume}{426}}, \bibinfo{pages}{55}
  (\bibinfo{year}{2003}{\natexlab{a}}).

\bibitem[{\citenamefont{Hur et~al.}(2004{\natexlab{a}})\citenamefont{Hur, Park,
  Sharma, Ahn, Guha, and Cheong}}]{Hur2004a}
\bibinfo{author}{\bibfnamefont{N.}~\bibnamefont{Hur}},
  \bibinfo{author}{\bibfnamefont{S.}~\bibnamefont{Park}},
  \bibinfo{author}{\bibfnamefont{P.~A.} \bibnamefont{Sharma}},
  \bibinfo{author}{\bibfnamefont{J.~S.} \bibnamefont{Ahn}},
  \bibinfo{author}{\bibfnamefont{S.}~\bibnamefont{Guha}}, \bibnamefont{and}
  \bibinfo{author}{\bibfnamefont{S.~W.} \bibnamefont{Cheong}},
  \bibinfo{journal}{Nature} \textbf{\bibinfo{volume}{429}},
  \bibinfo{pages}{392} (\bibinfo{year}{2004}{\natexlab{a}}).

\bibitem[{\citenamefont{Goto et~al.}(2004)\citenamefont{Goto, Kimura, Lawes,
  Ramirez, and Tokura}}]{Goto2004}
\bibinfo{author}{\bibfnamefont{T.}~\bibnamefont{Goto}},
  \bibinfo{author}{\bibfnamefont{T.}~\bibnamefont{Kimura}},
  \bibinfo{author}{\bibfnamefont{G.}~\bibnamefont{Lawes}},
  \bibinfo{author}{\bibfnamefont{A.~P.} \bibnamefont{Ramirez}},
  \bibnamefont{and} \bibinfo{author}{\bibfnamefont{Y.}~\bibnamefont{Tokura}},
  \bibinfo{journal}{Phys. Rev. Lett.} \textbf{\bibinfo{volume}{92}},
  \bibinfo{pages}{257201} (\bibinfo{year}{2004}).

\bibitem[{\citenamefont{Lottermoser et~al.}(2004)\citenamefont{Lottermoser,
  Lonkai, Amann, Hohlwein, Ihringer, and Fiebig}}]{Lottermoser2004}
\bibinfo{author}{\bibfnamefont{T.}~\bibnamefont{Lottermoser}},
  \bibinfo{author}{\bibfnamefont{T.}~\bibnamefont{Lonkai}},
  \bibinfo{author}{\bibfnamefont{U.}~\bibnamefont{Amann}},
  \bibinfo{author}{\bibfnamefont{D.}~\bibnamefont{Hohlwein}},
  \bibinfo{author}{\bibfnamefont{J.}~\bibnamefont{Ihringer}}, \bibnamefont{and}
  \bibinfo{author}{\bibfnamefont{M.}~\bibnamefont{Fiebig}},
  \bibinfo{journal}{Nature} \textbf{\bibinfo{volume}{430}},
  \bibinfo{pages}{541} (\bibinfo{year}{2004}).

\bibitem[{\citenamefont{Lorenz et~al.}(2004)\citenamefont{Lorenz, Wang, Sun,
  and Chu}}]{Lorenz2004}
\bibinfo{author}{\bibfnamefont{B.}~\bibnamefont{Lorenz}},
  \bibinfo{author}{\bibfnamefont{Y.~Q.} \bibnamefont{Wang}},
  \bibinfo{author}{\bibfnamefont{Y.~Y.} \bibnamefont{Sun}}, \bibnamefont{and}
  \bibinfo{author}{\bibfnamefont{C.~W.} \bibnamefont{Chu}},
  \bibinfo{journal}{Phys. Rev. B} \textbf{\bibinfo{volume}{70}},
  \bibinfo{pages}{212412} (\bibinfo{year}{2004}).

\bibitem[{\citenamefont{Higashiyama et~al.}(2004)\citenamefont{Higashiyama,
  Miyasaka, Kida, Arima, and Tokura}}]{Higashiyama2004}
\bibinfo{author}{\bibfnamefont{D.}~\bibnamefont{Higashiyama}},
  \bibinfo{author}{\bibfnamefont{S.}~\bibnamefont{Miyasaka}},
  \bibinfo{author}{\bibfnamefont{N.}~\bibnamefont{Kida}},
  \bibinfo{author}{\bibfnamefont{T.}~\bibnamefont{Arima}}, \bibnamefont{and}
  \bibinfo{author}{\bibfnamefont{Y.}~\bibnamefont{Tokura}},
  \bibinfo{journal}{Phys. Rev. B} \textbf{\bibinfo{volume}{70}},
  \bibinfo{pages}{174405} (\bibinfo{year}{2004}).

\bibitem[{\citenamefont{Hill}(2000)}]{Hill}
\bibinfo{author}{\bibfnamefont{N.~A.} \bibnamefont{Hill}}, \bibinfo{journal}{J.
  Phys. Chem. B} \textbf{\bibinfo{volume}{104}}, \bibinfo{pages}{6694}
  (\bibinfo{year}{2000}).

\bibitem[{\citenamefont{Fiebig}(2005)}]{Fiebig}
\bibinfo{author}{\bibfnamefont{M.}~\bibnamefont{Fiebig}}, \bibinfo{journal}{J.
  Phys. D} \textbf{\bibinfo{volume}{38}}, \bibinfo{pages}{R123}
  (\bibinfo{year}{2005}).

\bibitem[{\citenamefont{Khomskii}(2006)}]{Khomskii}
\bibinfo{author}{\bibfnamefont{D.~I.} \bibnamefont{Khomskii}},
  \bibinfo{journal}{J. Magn. Magn. Mater.} \textbf{\bibinfo{volume}{306}},
  \bibinfo{pages}{1} (\bibinfo{year}{2006}).

\bibitem[{\citenamefont{Cheong and Mostovoy}(2007)}]{Cheong2007}
\bibinfo{author}{\bibfnamefont{S.~W.} \bibnamefont{Cheong}} \bibnamefont{and}
  \bibinfo{author}{\bibfnamefont{M.}~\bibnamefont{Mostovoy}},
  \bibinfo{journal}{Nat. Mater.} \textbf{\bibinfo{volume}{6}},
  \bibinfo{pages}{13} (\bibinfo{year}{2007}).

\bibitem[{\citenamefont{Kimura et~al.}(2003{\natexlab{b}})\citenamefont{Kimura,
  Kawamoto, Yamada, Azuma, Takano, and Tokura}}]{Kimura2003}
\bibinfo{author}{\bibfnamefont{T.}~\bibnamefont{Kimura}},
  \bibinfo{author}{\bibfnamefont{S.}~\bibnamefont{Kawamoto}},
  \bibinfo{author}{\bibfnamefont{I.}~\bibnamefont{Yamada}},
  \bibinfo{author}{\bibfnamefont{M.}~\bibnamefont{Azuma}},
  \bibinfo{author}{\bibfnamefont{M.}~\bibnamefont{Takano}}, \bibnamefont{and}
  \bibinfo{author}{\bibfnamefont{Y.}~\bibnamefont{Tokura}},
  \bibinfo{journal}{Phys. Rev. B} \textbf{\bibinfo{volume}{67}},
  \bibinfo{pages}{180401} (\bibinfo{year}{2003}{\natexlab{b}}).

\bibitem[{\citenamefont{Kimura et~al.}(2005{\natexlab{a}})\citenamefont{Kimura,
  Lawes, Goto, Tokura, and Ramirez}}]{Kimura2005a}
\bibinfo{author}{\bibfnamefont{T.}~\bibnamefont{Kimura}},
  \bibinfo{author}{\bibfnamefont{G.}~\bibnamefont{Lawes}},
  \bibinfo{author}{\bibfnamefont{T.}~\bibnamefont{Goto}},
  \bibinfo{author}{\bibfnamefont{Y.}~\bibnamefont{Tokura}}, \bibnamefont{and}
  \bibinfo{author}{\bibfnamefont{A.~P.} \bibnamefont{Ramirez}},
  \bibinfo{journal}{Phys. Rev. B} \textbf{\bibinfo{volume}{71}},
  \bibinfo{pages}{224425} (\bibinfo{year}{2005}{\natexlab{a}}).

\bibitem[{\citenamefont{Hur et~al.}(2004{\natexlab{b}})\citenamefont{Hur, Park,
  Sharma, Guha, and Cheong}}]{guha}
\bibinfo{author}{\bibfnamefont{N.}~\bibnamefont{Hur}},
  \bibinfo{author}{\bibfnamefont{S.}~\bibnamefont{Park}},
  \bibinfo{author}{\bibfnamefont{P.~A.} \bibnamefont{Sharma}},
  \bibinfo{author}{\bibfnamefont{S.}~\bibnamefont{Guha}}, \bibnamefont{and}
  \bibinfo{author}{\bibfnamefont{S.-W.} \bibnamefont{Cheong}},
  \bibinfo{journal}{Phys. Rev. Lett.} \textbf{\bibinfo{volume}{93}},
  \bibinfo{pages}{107207} (\bibinfo{year}{2004}{\natexlab{b}}).

\bibitem[{\citenamefont{Kimura et~al.}(2006)\citenamefont{Kimura, Lashley, and
  Ramirez}}]{Kimura2006}
\bibinfo{author}{\bibfnamefont{T.}~\bibnamefont{Kimura}},
  \bibinfo{author}{\bibfnamefont{J.~C.} \bibnamefont{Lashley}},
  \bibnamefont{and} \bibinfo{author}{\bibfnamefont{A.~P.}
  \bibnamefont{Ramirez}}, \bibinfo{journal}{Phys. Rev. B}
  \textbf{\bibinfo{volume}{73}}, \bibinfo{pages}{220401}
  (\bibinfo{year}{2006}).

\bibitem[{\citenamefont{Yamasaki et~al.}(2006)\citenamefont{Yamasaki, Miyasaka,
  Kaneko, He, Arima, and Tokura}}]{Yamasaki2006}
\bibinfo{author}{\bibfnamefont{Y.}~\bibnamefont{Yamasaki}},
  \bibinfo{author}{\bibfnamefont{S.}~\bibnamefont{Miyasaka}},
  \bibinfo{author}{\bibfnamefont{Y.}~\bibnamefont{Kaneko}},
  \bibinfo{author}{\bibfnamefont{J.~P.} \bibnamefont{He}},
  \bibinfo{author}{\bibfnamefont{T.}~\bibnamefont{Arima}}, \bibnamefont{and}
  \bibinfo{author}{\bibfnamefont{Y.}~\bibnamefont{Tokura}},
  \bibinfo{journal}{Phys. Rev. Lett.} \textbf{\bibinfo{volume}{96}},
  \bibinfo{pages}{207204} (\bibinfo{year}{2006}).

\bibitem[{\citenamefont{Taniguchi et~al.}(2006)\citenamefont{Taniguchi, Abe,
  Takenobu, Iwasa, and Arima}}]{Taniguchi2006}
\bibinfo{author}{\bibfnamefont{K.}~\bibnamefont{Taniguchi}},
  \bibinfo{author}{\bibfnamefont{N.}~\bibnamefont{Abe}},
  \bibinfo{author}{\bibfnamefont{T.}~\bibnamefont{Takenobu}},
  \bibinfo{author}{\bibfnamefont{Y.}~\bibnamefont{Iwasa}}, \bibnamefont{and}
  \bibinfo{author}{\bibfnamefont{T.}~\bibnamefont{Arima}},
  \bibinfo{journal}{Phys. Rev. Lett.} \textbf{\bibinfo{volume}{97}},
  \bibinfo{pages}{097203} (\bibinfo{year}{2006}).

\bibitem[{\citenamefont{Lawes et~al.}(2005)\citenamefont{Lawes, Harris, Kimura,
  Rogado, Cava, Aharony, Entin-Wohlman, Yildirim, Kenzelmann, Broholm
  et~al.}}]{Lawes2005}
\bibinfo{author}{\bibfnamefont{G.}~\bibnamefont{Lawes}},
  \bibinfo{author}{\bibfnamefont{A.~B.} \bibnamefont{Harris}},
  \bibinfo{author}{\bibfnamefont{T.}~\bibnamefont{Kimura}},
  \bibinfo{author}{\bibfnamefont{N.}~\bibnamefont{Rogado}},
  \bibinfo{author}{\bibfnamefont{R.~J.} \bibnamefont{Cava}},
  \bibinfo{author}{\bibfnamefont{A.}~\bibnamefont{Aharony}},
  \bibinfo{author}{\bibfnamefont{O.}~\bibnamefont{Entin-Wohlman}},
  \bibinfo{author}{\bibfnamefont{T.}~\bibnamefont{Yildirim}},
  \bibinfo{author}{\bibfnamefont{M.}~\bibnamefont{Kenzelmann}},
  \bibinfo{author}{\bibfnamefont{C.}~\bibnamefont{Broholm}},
  \bibnamefont{et~al.}, \bibinfo{journal}{Phys. Rev. Lett.}
  \textbf{\bibinfo{volume}{95}}, \bibinfo{pages}{087205}
  (\bibinfo{year}{2005}).

\bibitem[{\citenamefont{Kimura et~al.}(2005{\natexlab{b}})\citenamefont{Kimura,
  Lawes, and Ramirez}}]{Kimura2005}
\bibinfo{author}{\bibfnamefont{T.}~\bibnamefont{Kimura}},
  \bibinfo{author}{\bibfnamefont{G.}~\bibnamefont{Lawes}}, \bibnamefont{and}
  \bibinfo{author}{\bibfnamefont{A.~P.} \bibnamefont{Ramirez}},
  \bibinfo{journal}{Phys. Rev. Lett.} \textbf{\bibinfo{volume}{94}},
  \bibinfo{pages}{137201} (\bibinfo{year}{2005}{\natexlab{b}}).

\bibitem[{com()}]{comm}
\bibinfo{note}{Private communications with Prof. S.W.Cheong. A correct
  identification of the crystal structure axes cannot be made unambiguously.}

\bibitem[{\citenamefont{Moskvin and Drechsler}(2008)}]{moskvin-2008}
\bibinfo{author}{\bibfnamefont{A.~S.} \bibnamefont{Moskvin}} \bibnamefont{and}
  \bibinfo{author}{\bibfnamefont{S.~L.} \bibnamefont{Drechsler}}
  (\bibinfo{year}{2008}), \bibinfo{note}{arXiv.org:0801.1102}.

\bibitem[{\citenamefont{Moskvin et~al.}(2008)\citenamefont{Moskvin, Panov, and
  Drechsler}}]{drechsler-2008}
\bibinfo{author}{\bibfnamefont{A.~S.} \bibnamefont{Moskvin}},
  \bibinfo{author}{\bibfnamefont{Y.~D.} \bibnamefont{Panov}}, \bibnamefont{and}
  \bibinfo{author}{\bibfnamefont{S.~L.} \bibnamefont{Drechsler}}
  (\bibinfo{year}{2008}), \bibinfo{note}{arXiv.org:0801.1975}.

\bibitem[{\citenamefont{Katsura et~al.}(2005)\citenamefont{Katsura, Nagaosa,
  and Balatsky}}]{Katsura2005}
\bibinfo{author}{\bibfnamefont{H.}~\bibnamefont{Katsura}},
  \bibinfo{author}{\bibfnamefont{N.}~\bibnamefont{Nagaosa}}, \bibnamefont{and}
  \bibinfo{author}{\bibfnamefont{A.~V.} \bibnamefont{Balatsky}},
  \bibinfo{journal}{Phys. Rev. Lett.} \textbf{\bibinfo{volume}{95}},
  \bibinfo{pages}{057205} (\bibinfo{year}{2005}).

\bibitem[{\citenamefont{Hu}(2008)}]{Hu2008}
\bibinfo{author}{\bibfnamefont{J.}~\bibnamefont{Hu}}, \bibinfo{journal}{Phys.
  Rev. Lett.} \textbf{\bibinfo{volume}{100}}, \bibinfo{pages}{077202}
  (\bibinfo{year}{2008}).

\bibitem[{\citenamefont{Sergienko and Dagotto}(2006)}]{Sergienko2006}
\bibinfo{author}{\bibfnamefont{I.~A.} \bibnamefont{Sergienko}}
  \bibnamefont{and} \bibinfo{author}{\bibfnamefont{E.}~\bibnamefont{Dagotto}},
  \bibinfo{journal}{Phys. Rev. B} \textbf{\bibinfo{volume}{73}},
  \bibinfo{pages}{094434} (\bibinfo{year}{2006}).

\bibitem[{\citenamefont{Mostovoy}(2006)}]{mostovoy}
\bibinfo{author}{\bibfnamefont{M.}~\bibnamefont{Mostovoy}},
  \bibinfo{journal}{Phys. Rev. Lett.} \textbf{\bibinfo{volume}{96}},
  \bibinfo{pages}{067601} (\bibinfo{year}{2006}).

\bibitem[{\citenamefont{Harris and Lawes}(2005)}]{Harris2005a}
\bibinfo{author}{\bibfnamefont{A.~B.} \bibnamefont{Harris}} \bibnamefont{and}
  \bibinfo{author}{\bibfnamefont{G.}~\bibnamefont{Lawes}},
  \bibinfo{journal}{cond-mat/0508617}  (\bibinfo{year}{2005}).

\bibitem[{\citenamefont{Choi et~al.}(2004)\citenamefont{Choi, Zvyagin, Cao, and
  Lemmens}}]{choi}
\bibinfo{author}{\bibfnamefont{K.-Y.} \bibnamefont{Choi}},
  \bibinfo{author}{\bibfnamefont{S.~A.} \bibnamefont{Zvyagin}},
  \bibinfo{author}{\bibfnamefont{G.}~\bibnamefont{Cao}}, \bibnamefont{and}
  \bibinfo{author}{\bibfnamefont{P.}~\bibnamefont{Lemmens}},
  \bibinfo{journal}{Phys. Rev. B} \textbf{\bibinfo{volume}{69}},
  \bibinfo{pages}{104421} (\bibinfo{year}{2004}).

\bibitem[{\citenamefont{Berger et~al.}(1991)\citenamefont{Berger, Meetsma, van
  Smaalen, and Sundberg}}]{Berger}
\bibinfo{author}{\bibfnamefont{R.}~\bibnamefont{Berger}},
  \bibinfo{author}{\bibfnamefont{A.}~\bibnamefont{Meetsma}},
  \bibinfo{author}{\bibfnamefont{S.}~\bibnamefont{van Smaalen}},
  \bibnamefont{and} \bibinfo{author}{\bibfnamefont{M.}~\bibnamefont{Sundberg}},
  \bibinfo{journal}{J. Less-Common Met.} \textbf{\bibinfo{volume}{175}},
  \bibinfo{pages}{119} (\bibinfo{year}{1991}).

\bibitem[{\citenamefont{Zvyagin et~al.}(2002)\citenamefont{Zvyagin, Cao, Xin,
  McCall, Caldwell, Moulton, Brunel, Angerhofer, and Crow}}]{Zvyagin}
\bibinfo{author}{\bibfnamefont{S.}~\bibnamefont{Zvyagin}},
  \bibinfo{author}{\bibfnamefont{G.}~\bibnamefont{Cao}},
  \bibinfo{author}{\bibfnamefont{Y.}~\bibnamefont{Xin}},
  \bibinfo{author}{\bibfnamefont{S.}~\bibnamefont{McCall}},
  \bibinfo{author}{\bibfnamefont{T.}~\bibnamefont{Caldwell}},
  \bibinfo{author}{\bibfnamefont{W.}~\bibnamefont{Moulton}},
  \bibinfo{author}{\bibfnamefont{L.-C.} \bibnamefont{Brunel}},
  \bibinfo{author}{\bibfnamefont{A.}~\bibnamefont{Angerhofer}},
  \bibnamefont{and} \bibinfo{author}{\bibfnamefont{J.~E.} \bibnamefont{Crow}},
  \bibinfo{journal}{Phys. Rev. B} \textbf{\bibinfo{volume}{66}},
  \bibinfo{pages}{064424} (\bibinfo{year}{2002}).

\bibitem[{\citenamefont{Roessli et~al.}(2001)\citenamefont{Roessli, Staubb,
  Amatoc, Herlachc, Pattisond, Sablinae, and Petrakovskiie}}]{Roessli}
\bibinfo{author}{\bibfnamefont{B.}~\bibnamefont{Roessli}},
  \bibinfo{author}{\bibfnamefont{U.}~\bibnamefont{Staubb}},
  \bibinfo{author}{\bibfnamefont{A.}~\bibnamefont{Amatoc}},
  \bibinfo{author}{\bibfnamefont{D.}~\bibnamefont{Herlachc}},
  \bibinfo{author}{\bibfnamefont{P.}~\bibnamefont{Pattisond}},
  \bibinfo{author}{\bibfnamefont{K.}~\bibnamefont{Sablinae}}, \bibnamefont{and}
  \bibinfo{author}{\bibfnamefont{G.~A.} \bibnamefont{Petrakovskiie}},
  \bibinfo{journal}{Physica B} \textbf{\bibinfo{volume}{296}},
  \bibinfo{pages}{306} (\bibinfo{year}{2001}).

\bibitem[{\citenamefont{Masuda et~al.}(2004)\citenamefont{Masuda, Zheludev,
  Bush, Markina, and Vasiliev}}]{masuda}
\bibinfo{author}{\bibfnamefont{T.}~\bibnamefont{Masuda}},
  \bibinfo{author}{\bibfnamefont{A.}~\bibnamefont{Zheludev}},
  \bibinfo{author}{\bibfnamefont{A.}~\bibnamefont{Bush}},
  \bibinfo{author}{\bibfnamefont{M.}~\bibnamefont{Markina}}, \bibnamefont{and}
  \bibinfo{author}{\bibfnamefont{A.}~\bibnamefont{Vasiliev}},
  \bibinfo{journal}{Phys. Rev. Lett.} \textbf{\bibinfo{volume}{92}},
  \bibinfo{pages}{177201} (\bibinfo{year}{2004}).

\bibitem[{out()}]{outofplane}
\bibinfo{note}{While the neutron diffraction experiments \cite{masuda} conclude
  that rotating spins lie in one plane ($bc$), there is independent NMR studies
  \cite{gippius} which show an out-of-plane component.}

\bibitem[{\citenamefont{Capogna et~al.}(2005)\citenamefont{Capogna, Mayr,
  Horsch, Raichle, Kremer, Sofin, Maljuk, Jansen, and Keimer}}]{capogna}
\bibinfo{author}{\bibfnamefont{L.}~\bibnamefont{Capogna}},
  \bibinfo{author}{\bibfnamefont{M.}~\bibnamefont{Mayr}},
  \bibinfo{author}{\bibfnamefont{P.}~\bibnamefont{Horsch}},
  \bibinfo{author}{\bibfnamefont{M.}~\bibnamefont{Raichle}},
  \bibinfo{author}{\bibfnamefont{R.~K.} \bibnamefont{Kremer}},
  \bibinfo{author}{\bibfnamefont{M.}~\bibnamefont{Sofin}},
  \bibinfo{author}{\bibfnamefont{A.}~\bibnamefont{Maljuk}},
  \bibinfo{author}{\bibfnamefont{M.}~\bibnamefont{Jansen}}, \bibnamefont{and}
  \bibinfo{author}{\bibfnamefont{B.}~\bibnamefont{Keimer}},
  \bibinfo{journal}{Phys. Rev. B} \textbf{\bibinfo{volume}{71}},
  \bibinfo{pages}{140402} (\bibinfo{year}{2005}).

\bibitem[{\citenamefont{Choi et~al.}(2006)\citenamefont{Choi, Gnezdilov,
  Lemmens, Capogna, Johnson, Sofin, Maljuk, Jansen, and Keimer}}]{Gnezdilov}
\bibinfo{author}{\bibfnamefont{K.-Y.} \bibnamefont{Choi}},
  \bibinfo{author}{\bibfnamefont{V.~P.} \bibnamefont{Gnezdilov}},
  \bibinfo{author}{\bibfnamefont{P.}~\bibnamefont{Lemmens}},
  \bibinfo{author}{\bibfnamefont{L.}~\bibnamefont{Capogna}},
  \bibinfo{author}{\bibfnamefont{M.~R.} \bibnamefont{Johnson}},
  \bibinfo{author}{\bibfnamefont{M.}~\bibnamefont{Sofin}},
  \bibinfo{author}{\bibfnamefont{A.}~\bibnamefont{Maljuk}},
  \bibinfo{author}{\bibfnamefont{M.}~\bibnamefont{Jansen}}, \bibnamefont{and}
  \bibinfo{author}{\bibfnamefont{B.}~\bibnamefont{Keimer}},
  \bibinfo{journal}{Phys. Rev. B} \textbf{\bibinfo{volume}{73}},
  \bibinfo{pages}{094409} (\bibinfo{year}{2006}).

\bibitem[{\citenamefont{Sushkov et~al.}(2007)\citenamefont{Sushkov, Aguilar,
  Park, Cheong, and Drew}}]{Sushkov2007}
\bibinfo{author}{\bibfnamefont{A.~B.} \bibnamefont{Sushkov}},
  \bibinfo{author}{\bibfnamefont{R.~V.} \bibnamefont{Aguilar}},
  \bibinfo{author}{\bibfnamefont{S.}~\bibnamefont{Park}},
  \bibinfo{author}{\bibfnamefont{S.-W.} \bibnamefont{Cheong}},
  \bibnamefont{and} \bibinfo{author}{\bibfnamefont{H.~D.} \bibnamefont{Drew}},
  \bibinfo{journal}{Phys. Rev. Lett.} \textbf{\bibinfo{volume}{98}},
  \bibinfo{pages}{027202} (\bibinfo{year}{2007}).

\bibitem[{\citenamefont{Fang and Hu}(2008)}]{chen:epl}
\bibinfo{author}{\bibfnamefont{C.}~\bibnamefont{Fang}} \bibnamefont{and}
  \bibinfo{author}{\bibfnamefont{J.}~\bibnamefont{Hu}} (\bibinfo{year}{2008}),
  \bibinfo{note}{"An effective model of magnetoelectricity in multiferroics
  RMn(2)O(5), to appear in Europhys. Lett."}.

\bibitem[{\citenamefont{Gippius et~al.}(2004)\citenamefont{Gippius, Morozova,
  Moskvin, Zalessky, Bush, Baenitz, Rosner, and Drechsler}}]{gippius}
\bibinfo{author}{\bibfnamefont{A.~A.} \bibnamefont{Gippius}},
  \bibinfo{author}{\bibfnamefont{E.~N.} \bibnamefont{Morozova}},
  \bibinfo{author}{\bibfnamefont{A.~S.} \bibnamefont{Moskvin}},
  \bibinfo{author}{\bibfnamefont{A.~V.} \bibnamefont{Zalessky}},
  \bibinfo{author}{\bibfnamefont{A.~A.} \bibnamefont{Bush}},
  \bibinfo{author}{\bibfnamefont{M.}~\bibnamefont{Baenitz}},
  \bibinfo{author}{\bibfnamefont{H.}~\bibnamefont{Rosner}}, \bibnamefont{and}
  \bibinfo{author}{\bibfnamefont{S.-L.} \bibnamefont{Drechsler}},
  \bibinfo{journal}{Phys. Rev. B} \textbf{\bibinfo{volume}{70}},
  \bibinfo{pages}{020406} (\bibinfo{year}{2004}).

\end{thebibliography}
\end{document}